\documentclass[prd,reprint,nofootinbib,notitlepage,aps,tightenlines,
preprintnumbers,amsmath,amssymb,showpacs,superscriptaddress]{revtex4-1}

\usepackage[hyperindex,breaklinks]{hyperref}
\usepackage{graphicx,epstopdf,amsmath,amsfonts,amssymb,color,comment,cancel,subfigure}

\usepackage{tikz}
\def\checkmark{\tikz\fill[scale=0.4](0,.35) -- (.25,0) -- (1,.7) -- (.25,.15) -- cycle;} 

\usepackage{chngcntr}
\counterwithin{paragraph}{subsection}

\def\be{\begin{equation}}
\def\ee{\end{equation}}
\def\ba{\begin{eqnarray}}
\def\ea{\end{eqnarray}}
\def\ge{\mathrel{\raise.3ex\hbox{$>$\kern-.75em\lower1ex\hbox{$\sim$}}}}
\def\la{\mathrel{\raise.3ex\hbox{$<$\kern-.75em\lower1ex\hbox{$\sim$}}}}

\def\simgt{\mathrel{\raise.3ex\hbox{$>$\kern-.75em\lower1ex\hbox{$\sim$}}}}
\def\simlt{\mathrel{\raise.3ex\hbox{$<$\kern-.75em\lower1ex\hbox{$\sim$}}}}

\newcommand{\nc}{\newcommand}

\nc{\gone}{\bar g_{\pi NN}^{(1)}}
\nc{\gzero}{\bar g_{\pi NN}^{(0)}}
\nc{\al}{\alpha}
\nc{\ga}{\gamma}
\nc{\de}{\delta}
\nc{\ep}{\epsilon}
\nc{\ze}{\zeta}
\nc{\et}{\eta}
\nc{\ka}{\kappa}
\nc{\rh}{\rho}
\nc{\si}{\sigma}
\nc{\ta}{\tau}
\nc{\up}{\upsilon}
\nc{\ph}{\phi}
\nc{\ch}{\chi}
\nc{\ps}{\psi}
\nc{\om}{\omega}
\nc{\Ga}{\Gamma}
\nc{\De}{\Delta}
\nc{\La}{\Lambda}
\nc{\Si}{\Sigma}
\nc{\Up}{\Upsilon}
\nc{\Ph}{\Phi}
\nc{\Ps}{\Psi}
\nc{\Om}{\Omega}
\nc{\ptl}{\partial}
\nc{\del}{\nabla}
\nc{\ov}{\overline}
\nc{\newcaption}[1]{\centerline{\parbox{15cm}{\caption{#1}}}}
\nc{\us}{U(1)$_S$}
\nc{\Rg}{$R_{\gamma\gamma}$}
\nc{\bbnu}{\beta\beta_{0\nu}}

\def\beq{\begin{equation}}
\def\eeq{\end{equation}}
\def\bmat{\begin{displaymath}}
\def\emat{\end{displaymath}}
\def\bear{\begin{eqnarray}}
\def\eear{\end{eqnarray}}
\def\ba{\begin{eqnarray}}
\def\ea{\end{eqnarray}}
\def\bery{\begin{array}}
\def\ery{\end{array}}
\def\bit{\begin{itemize}}
\def\eit{\end{itemize}}
\def\ben{\begin{enumerate}}
\def\een{\end{enumerate}}
\def\btab{\begin{tabular}}
\def\etab{\end{tabular}}
\def\btbl{\begin{table}}
\def\etbl{\end{table}}
\def\bfig{\begin{figure}[htb]}
\def\efig{\end{figure}}
\def\bpic{\begin{picture}}
\def\epic{\end{picture}}

\def\ga{\mathrel{\raise.3ex\hbox{$>$\kern-.75em\lower1ex\hbox{$\sim$}}}}
\def\la{\mathrel{\raise.3ex\hbox{$<$\kern-.75em\lower1ex\hbox{$\sim$}}}}
\def\gappeq{\mathrel{\rlap {\raise.5ex\hbox{$>$}}
{\lower.5ex\hbox{$\sim$}}}}
\def\lappeq{\mathrel{\rlap{\raise.5ex\hbox{$<$}}
{\lower.5ex\hbox{$\sim$}}}}

\def\gyr{{\rm \, G\kern-0.125em yr}}
\def\mev{{\rm \, Me\kern-0.125em V}}
\def\gev{{\rm \, Ge\kern-0.125em V}}
\def\tev{{\rm \, Te\kern-0.125em V}}

\def\lsim{\mathrel{\rlap{\lower4pt\hbox{\hskip1pt$\sim$}}
    \raise1pt\hbox{$<$}}}                % less than or approx. symbol
\def\gsim{\mathrel{\rlap{\lower4pt\hbox{\hskip1pt$\sim$}}
    \raise1pt\hbox{$>$}}}                % greater than or approx. symbol

\newcommand{\Leq}{\leqslant}

\definecolor{jazzberryjam}{rgb}{0.65, 0.04, 0.37}        % http://latexcolor.com/

\begin{document}
 
\title{Sensitivity to light weakly-coupled new physics at the precision frontier}
%Probing light new physics with precision frontier measurements}

\author{Matthias Le Dall}
\affiliation{Department of Physics and Astronomy, University of Victoria, 
Victoria, BC V8P 5C2, Canada}

\author{Maxim Pospelov}
\affiliation{Department of Physics and Astronomy, University of Victoria, 
Victoria, BC V8P 5C2, Canada}
\affiliation{Perimeter Institute for Theoretical Physics, Waterloo, ON N2J 2W9, 
Canada}

\author{Adam Ritz}
\affiliation{Department of Physics and Astronomy, University of Victoria, 
Victoria, BC V8P 5C2, Canada}

\date{May 2015}

\begin{abstract}

Precision measurements of rare particle physics phenomena (flavor oscillations and decays, electric dipole moments, etc.) are often 
sensitive to the effects of new physics encoded in higher-dimensional operators with Wilson coefficients given by ${\rm C}/(\Lambda_{\rm NP})^n$, 
where C is dimensionless, $n\geq 1$, and $\Lambda_{\rm NP}$ is an energy scale. Many extensions of the Standard Model predict
that $\Lambda_{\rm NP} $ should be at the electroweak scale or above, and the search for new short-distance physics
is often stated as the primary goal of experiments at the precision frontier. In rather general terms, we investigate the alternative 
possibility: ${\rm C} \ll 1$, and $\Lambda_{\rm NP} \ll m_W$,  to identify classes of precision measurements sensitive to light new physics 
(hidden sectors) that do not require an ultraviolet completion with additional states at or above the 
electroweak scale.  We find that hadronic electric dipole moments, lepton number and flavor violation, non-universality, as well as lepton $g-2$ can be induced at interesting 
levels by hidden sectors with light degrees of freedom. In contrast, many hadronic flavor- and baryon number-violating observables, and precision probes of charged currents, 
typically require new physics with $\Lambda_{\rm NP} \ga m_W$. Among the leptonic observables, we find that a non-zero electron electric dipole moment near the current level of 
sensitivity would point to the existence of new physics at or above the electroweak scale. 

\end{abstract}
\maketitle

\section{Introduction}
\label{sec:intro}

\begin{figure}[t!]
\centering
\vspace*{-0.5cm}
\includegraphics[width=8cm]{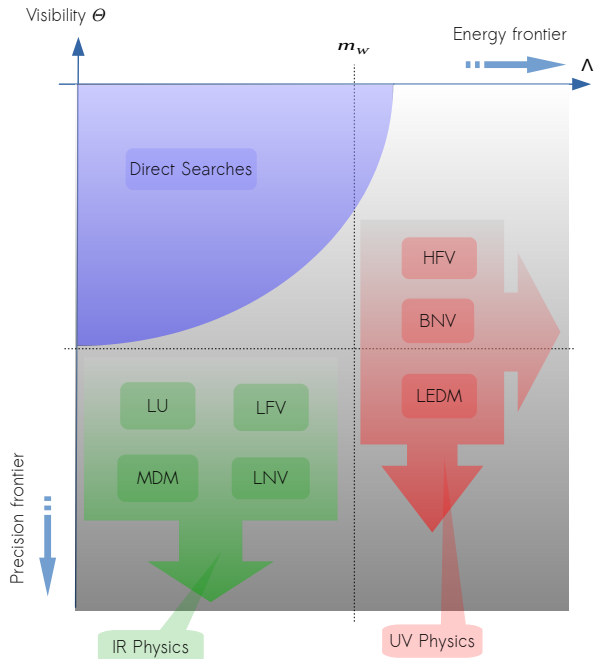}
\vspace*{-1.15cm}
\caption{A schematic view of the parameter space of mass scale vs coupling for physics beyond the SM. The horizontal axis represents the mass (or energy) scale, whereas the 
vertical scale shows the \textit{visibility} of the model, in terms of the coupling to the SM. The blue visible area is accessible through direct searches. While new high energy 
physics can contribute to all precision observables, as discussed in this paper there are interesting classes of observables that are also sensitive to low-mass new physics. These 
are shown in green, and test lepton universality (LU), lepton flavor violation (LFV), lepton $g-2$ (MDM), and lepton number violation (LNV). However, observables in the hadronic 
sector in red, e.g. hadronic flavor violation (HFV) and baryon number violation (BNV), and also lepton electric dipole moments (LEDMs) generally require some new high scale 
physics. The arrows indicate the pressure imposed on models through increasing experimental sensitivity.}
\label{fig: theory landscape}
\end{figure}

Accelerator-based particle physics has the  goal of probing the shortest distance scales directly, by colliding particles and their constituents at high energies. 
Thus far, all high energy data is well described by the Standard Model (SM) of particles and fields, with the last missing element, the Higgs boson, identified recently 
\cite{Aad:2012tfa,Chatrchyan:2012ufa}. Considerable attention is therefore focussed on the search for `new physics' (NP) that may complement the SM by addressing some of its 
shortcomings. However, the most prominent empirical evidence for new physics, associated for example with neutrino mass and dark matter, does not necessarily point to an origin at 
shorter distance scales. 

Fortunately, experiments at the energy frontier are not the only tools available to probe NP; they are supplemented by searches at the 
precision (and intensity) frontier (see {\em e.g.}~\cite{Raidal:2008jk}). Precision observables, particularly those that probe violations of exact or approximate symmetries of the 
Standard Model such as $CP$ and flavor, play an important role in the search for new physics \cite{Pospelov:2005pr, Engel:2013lsa, deGouvea:2013zba, Gedalia:2010rj}. Their reach in 
energy scale, through  loop-induced corrections from new UV physics, can often extend well beyond the direct reach of high energy colliders. However, measurements at low energies 
may be sensitive not only to NP corrections coming from the short distances, but also to NP at longer distances (lower mass) with extremely weak coupling to the SM.  It is 
therefore prudent to ask for which precision observables can measured deviations from SM predictions {\it unambiguously} be identified with short-distance NP at the electroweak 
(EW) scale or above? Alternatively, one can ask when such deviations might also admit an interpretation in terms of  
new low-scale hidden sector degrees of freedom. This is the question we will address in this paper.

The sensitivity of any constraint on new physics is determined on one hand by the precision of the measurement in question, and on the other by the accuracy and precision of any SM 
calculations required to disentangle background contributions. If the effective Lagrangian is schematically written in the form ${\cal L} = {\cal L}_{\rm SM}+ {\cal L}_{\rm NP}$, 
the possibility of discovery relies on being able to reliably bound the NP contribution to the observable away from zero. The natural tendency to interpret results in terms of 
operators in ${\cal L}_{\rm NP}$ induced by ultraviolet NP can be problematic, as ${\cal L}_{\rm NP}$ can in general also receive contributions from 
light weakly-coupled degrees of freedom. This dilemma is nicely illustrated by the theoretical interpretation of a NP discovery that has already occurred, namely the observation of 
neutrino flavor oscillations. The experimental results are most straightforwardly interpreted in terms of the masses and mixing of the light active neutrino species 
\cite{Maki:1962mu,Pontecorvo:1967fh}. However, as is well known, there are a number of possible explanations for their origin. These include a short-distance explanation in terms 
of the dimension-five Weinberg operator \cite{Weinberg:1979sa}, ${\cal L}_{\rm NP} \propto (HL)(HL)/\Lambda_{\rm UV}$ with $\Lambda_{\rm UV} \gg \langle H\rangle$,
which generates neutrino masses scaling as $\langle H\rangle^2/\Lambda_{\rm UV}$. There are also a variety of different UV completions for this operator, with and without heavy 
right-handed neutrino states, present throughout the theory literature. While this interpretation is certainly valid, there is also the possibility of interpreting 
neutrino mass as a consequence of very light states $N$, with $m_{N} \ll m_W$ and the quantum numbers of right-handed neutrinos \cite{Asaka:2005an, 
Akhmedov:1998qx,Asaka:2005pn,Petraki:2007gq,deGouvea:2005er,Kusenko:2009up}. Such states would typically be very weakly coupled to the SM, thus escaping direct detection. The most 
prominent model in this class is the simple three-generation extension of the SM with $N$ states that allow Dirac masses for the active neutrinos. Thus we see that neutrino 
oscillations can be interpreted as the result of UV or IR new physics (or both).

In this paper we  scrutinize several  classes of precision frontier measurements, and confront them with the possibility of 
NP confined {\it solely} to low energy (sub-EW) scales.  Our goal is to find specific examples of light NP that can induce 
$CP$-violation, cause deviations from calculated values of $g-2$, or lead to flavor changing effects. One condition we set on the classes of such models 
is the absence of any additional physics at or above the EW scale. In other words, we shall focus on UV complete models of light NP. 
Given the wealth of particle physics data, only very specific classes of light NP models can still be hidden below the weak scale. In Section~2, we describe this classification of 
infrared new physics scenarios in more detail. Then in Section~3 we  discuss a number of different leptonic and hadronic observables, and explore simple new physics scenarios which 
provide a possible interpretation of any deviation in precision measurements. We summarize the analysis in Section~4. A schematic overview of the results is presented in 
Fig.~\ref{fig: theory landscape}, which illustrates the classes of (primarily leptonic) observables that can naturally be interpreted in terms of light UV-complete new physics.

\section{UV and IR new physics}

A simple characterization of UV/IR new physics scenarios follows by making the division at the electroweak scale, so that the chiral electroweak $SU(2)_L\times U(1)_Y$ 
structure is maintained,
\be
 {\cal L}_{\rm NP} ={\cal L}_{\rm UV} + {\cal L}_{\rm IR}.
\ee
New UV physics can then universally be described at the EW scale by a series of higher dimensional operators constructed from SM degrees of freedom,
\be
 {\cal L}_{\rm UV} = \sum_{d\geq 5} \frac{1}{\La_{\rm UV}^{d-4}} {\cal O}_{d}.
\ee
Maintaining SM gauge invariance explicitly, we demand that ${\cal O}_d$ can be written in an $SU(2)_L\times U(1)_Y$ 
covariant form. The lowest dimension $d=5$ includes only $LHLH$-type operators, which contribute to neutrino mass. 
The number of operators grows rapidly at $d=6$ and above \cite{Weinberg:1980bf, *Weldon:1980gi, *Buchmuller:1985jz, *Babu:2001ex, *deGouvea:2007xp, *Bonnet:2009ej, 
*Grzadkowski:2010es, *delAguila:2012nu, *Angel:2012ug, *Babu:2012iv, *Babu:2012vb, *Chalons:2013mya, *Lehman:2014jma}. We impose no 
restrictions on these operators, other than that $\La_{\rm UV} \gg m_Z$, so that they can consistently be written in $SU(2)_L\times U(1)_Y$ covariant form. Unless these new 
operators violate some of the well-tested exact or approximate 
discrete symmetries of the SM, $\La_{\rm UV}$ can be taken fairly close to the EW scale.  It is important to notice that the new states appearing at $\La_{\rm UV}$ could be 
charged under any of the SM gauge groups, and some of the most stringent constraints in cases where no specific symmetries are violated now come from the LHC.

In comparison, new IR physics is rather more constrained. A convenient categorization of light NP scenarios can be constructed as follows:

\begin{enumerate}
\item[A.] {\it Portals}:  Neutral hidden sectors, with operators of dimension $d\leq 4$, can couple through a restricted set of renormalizable 
interaction channels, the vector, Higgs and neutrino portals (see e.g. \cite{Essig:2013lka}). Such models of light new physics are fully UV complete without any additional charged 
states.
\item[B.] {\it Anomaly free (neutral)}: Light hidden sectors can also be charged under anomaly-free combinations of SM symmetries. For those combinations, such as $B-L$ or $L_\mu - 
L_\tau$, that do not involve individual quark flavors, additional (light and neutral) Higgs fields may be necessary to retain a viable mass spectrum, but these extra states can be 
SM-neutral. Therefore, these scenarios also fall into the category of UV-complete and gauge-neutral hidden sectors.
\item[C.] {\it Anomaly free (charged)}: Light hidden sectors charged under anomaly-free, but quark-flavor non-universal, symmetries such as $Q_{f_1}-Q_{f_2}$
require new charged Higgs states to restore the mass spectrum. Thus, these new physics scenarios generally require charged states at or above the EW scale.
\item[D.] {\it Anomalous}: Light hidden sectors charged under anomalous SM symmetries, such as $B$ or $L$, necessarily require additional (heavy) charged states at or above the EW 
scale, and so again do not fall into the category of IR new physics scenarios considered here. Indeed, as emphasized for example by Preskill \cite{Preskill:1990fr}, from the low 
energy perspective, anomalous theories are phenomenologically analogous to UV new physics scenarios with a specific UV cutoff.
\end{enumerate}

Based on this categorization, we will limit our attention to cases A and B, namely those which do not require new charged states at or above the EW scale for consistency. Thus 
we construct our model examples according to the following rules:
\begin{itemize}
\item The dimensionality of operators in the IR sector is restricted to $d\leq 4$, as a necessary condition for 
UV completeness.  

\item The IR sector cannot contain new SM-charged states. (Otherwise, such states will have to be close to or above the EW scale modulo some 
exceptional cases where masses as low as $\sim 60$ GeV may still be viable \cite{Batell:2013bka}). New charged states fall into the category 
of NP at the EW scale, and form part of ${\cal L}_{\rm UV}$.

\item The gauge extensions of the SM are restricted to anomaly-free combinations, which is also 
a generic requirement of UV completeness \cite{Preskill:1990fr}. 

\item We shall not question naturalness of possible mass hierarchies, $m_{\rm IR} \ll m_W$, 
and will take them as given. 
\end{itemize}
 
The simplest type of neutral hidden sector (case A) requires new scalars $S_i$, neutral fermions $N_i$ and/or new $U(1)$ gauge boson(s) $V_\mu$ \cite{Holdom:1985ag,*Foot:1991kb, 
*Foot:1991bp,*Pospelov:2007mp,*Batell:2009di}.  The most economical renormalizable portal interactions for these states can be written in the form
\be
\label{portal}
 {\cal L}_{\rm IR} = \kappa B^{\mu\nu}V_{\mu\nu} - H^\dagger H (AS + \lambda S^2) - Y_N LHN + {\cal L}_{\rm hid},
\ee
and can trivially be  generalized to multiple new fields and to a charged version of $S$, $S^2\to |S|^2$. Once coupled to the SM through these channels, the IR hidden sector can be 
almost arbitrarily complicated. $S$ and $N$ can couple to a complex hidden sector involving dark abelian or non-abelian gauge groups, possibly with additional scalar or fermion 
states charged under those hidden gauge groups. The full hidden sector Lagrangian simply needs to comply with the conditions above. 
The portal interactions in (\ref{portal}) are complete under the assumption that the SM is strictly neutral under the extra $U(1)$. However, this is unnecessarily restrictive. 
Light NP models (in case B) may also include non-anomalous gauged versions of global symmetries such as $B-L$ and $L_i-L_j$ etc, 
where SM fields receive charges under the new $U(1)$. 

It is also important to discuss some examples of theories that {\em do not} satisfy the above criteria. 
For example, a light pseudoscalar $a$ coupled via the axion portal to a SM fermion $\psi$, 
$\frac{1}{f_a}\partial_\mu a\,  \bar \psi \gamma^\mu\gamma_5 \psi$,
clearly requires UV completion at some high energy scale $\sim f_a$. Interestingly, a light scalar directly coupled 
to the scalar fermion density, $S \bar \psi \psi$, is allowed, provided that this coupling descends from the 
Higgs portal $ASH^\dagger H$, once the heavy SM Higgs particle is integrated out. This means, of course, that the 
ratio of the effective Yukawa couplings of $S$ to $\psi$ will obey the same relations as in the SM, and any deviations from this pattern would 
imply the existence of new Higgs doublets charged under the SM, and hence some new physics at or above the EW scale. 

\begin{table}
\begin{tabular}{|c|c|c|c|c|}
\hline
 Observable & (A,B) Portals & (C,D) UV-incomplete \\ \hline
 LFV & \checkmark & \checkmark  \\
 LU &  \checkmark & \checkmark \\
 $(g-2)_{l}$ &  \checkmark &  \checkmark \\
 LNV &  \checkmark & \checkmark \\
 LEDMs &   & \checkmark \\
 HFV & & \checkmark \\
 BNV & & \checkmark \\ \hline
\end{tabular}
\caption{Observables sensitive to the distinct classes of light new physics models discussed in Section 2.}
\end{table}

We turn in the next section to discuss a range of precision observables, and seek to determine which of them can receive significant contributions from IR new physics. Table~1 
summarizes the results from the next section, and refines the schematic classification of Fig.~1 according to the categorization A--D of new physics models introduced above.

\section{Precision observables}

\subsection{Lepton anomalous magnetic moments}

The anomalous magnetic moments of the electron and the muon represent observables \cite{Odom:2006zz,Bennett:2006fi}
where the SM contribution can be evaluated to high accuracy. For electrons, the sensitivity to NP depends on an independent determination of the electromagnetic fine structure 
constant. Currently, $g-2$ of the electron  (and related measurements) probe NP contributions at the level 
$\Delta a_e({\rm NP}) < 1.64\times 10^{-12}$ (see {\em e.g.} \cite{Davoudiasl:2012ig}), 
whereas $g-2$ of the muon famously exhibits a roughly $3.5 \sigma$ discrepancy \cite{Bennett:2006fi} 
between the measurement and the SM prediction, with the central value giving $\Delta a_\mu \simeq 3\times 10^{-9}$. 

It is tempting to interpret this discrepancy as a consequence of NP that adds a positive contribution 
to the predicted SM value. While many UV interpretations exist
(see {\em e.g.} \cite{Czarnecki:2001pv} ), $\Delta a_\mu ({\rm NP}) $ can just as easily result from
one-loop contributions of light particles. At the 
effective Lagrangian level, both $g' V_\nu \bar \mu \gamma^\nu \mu$ and $\lambda'  S \bar \mu  \mu$ 
can supply the requisite correction,
\be
g', \lambda' \sim 10^{-3}~{\rm with}~ m_{V(S)} \la m_\mu ~\Longrightarrow ~ \Delta a_\mu \sim 10^{-9}.
\ee
This fact is well-appreciated in the literature \cite{Pospelov:2008zw,Fayet:2007ua,Gninenko:2001hx}. 

The vector model has UV completions involving a kinetically mixed vector, or alternatively a symmetry 
based on gauged muon number, such as $L_\mu-L_\tau$. Dedicated searches for `dark photons' 
\cite{Essig:2013lka} have now placed a number of restrictions on the parameter space of this model. At this point, the kinetically 
mixed vector option to explain $\Delta a_\mu \simeq 3\times 10^{-9}$ is essentially ruled out through direct production experiments
{\em assuming} $V$ decays back primarily to SM states \cite{Lees:2014xha,Goudzovski:2014rwa}. Moreover, the alternative option of vectors decaying to light dark matter is 
also significantly 
constrained \cite{Batell:2014yra,Batell:2014mga,Kahn:2014sra}. On the other hand, a model with mutliple (cascading) decays of $V$ into the hidden 
sector can be ruled out only via missing energy signatures, and up to now, significant parameter space is still open for $\Delta a_\mu \sim O(10^{-9})$.
The $L_\mu-L_\tau$ explanation is even less constrained, with only trident neutrino production providing 
an adequate level of sensitivity \cite{Altmannshofer:2014pba,Altmannshofer:2014cfa}. 
We conclude that there are multiple IR models of NP that can lead to the observable 
shifts in $g-2$, while at the same time escaping direct detection constraints. 

In contrast, models based on scalar particles do not provide large shifts to $\Delta a_\mu$. For example, 
UV completion via Higgs mixing would imply
\be
\lambda' \simeq \frac{ A m_\mu}{m_h^2} \ll 10^{-3} ,
\ee
as other constraints on the model force $A$ to be much smaller than the weak scale, $A\ll m_h$.
 Thus, larger values of $\lambda'$ would require additional NP to appear at the weak scale.

\subsection{Lepton flavor violation and universality}

We now turn to leptonic flavor-violating observables. In analyzing IR new physics scenarios, it will be convenient to have in mind a specific hidden sector coupled through the  
neutrino portal. In particular, to the three left-handed active neutrinos $\nu_l$, $l=e,\mu,\tau$, we add corresponding right-handed neutrinos $N_{R}$, plus a number of extra 
singlet fermion states $N_{S}$. In our search for light NP models able to induce appreciable deviations of precision measurements in the lepton sector, it will prove advantageous 
to look in detail at a model that uses an inverse seesaw scheme for neutrino masses \cite{Mohapatra:1986bd, *Mohapatra:1986aw, *GonzalezGarcia:1988rw, *Fraser:2014yha}, 
\beq
 -\mathcal{L}_{\nu}\supset\left(\nu_L\quad N_R\quad N_S\right)\begin{pmatrix}
                    0 && m_D && 0\\
                    m_D && 0 && M_D\\
                    0 && M_D && \epsilon
                   \end{pmatrix}\begin{pmatrix}
                                 \nu_L\\ N_R\\ N_S
                                \end{pmatrix},
\eeq
in the regime $\epsilon\ll m_D,M_D$. The Dirac mass terms $m_D,M_D$ are matrices, but in the simplest example of only one active flavor, one right-handed neutrino and one  
additional singlet, this model predicts one light and two heavy mass eigenstates, $m_\nu,M_\pm$ respectively, given by
\beq
m_\nu\sim\frac{m_D^2}{R^2}\epsilon +\mathcal{O}(\epsilon^2),\quad M_{\pm}\sim R\pm\frac{M_D^2}{2R^2}\epsilon +\mathcal{O}(\epsilon^2),
\eeq
with $R^2=m_D^2+M_D^2$. In order to accommodate the light neutrino mass spectrum, we choose $m_D\lesssim M_D$, and $\epsilon\gtrsim m_\nu$. To lowest order, the unitary matrix $U$ 
that 
transforms the mass eigenstates into the flavor eigenstates takes the form,
\beq
\begin{pmatrix}\nu_L\\ N_R\\ N_S\end{pmatrix}=U\begin{pmatrix}\nu_i\\ N_+\\ N_-\end{pmatrix},
\quad
 U\approx\begin{pmatrix}
    \frac{M_D}{R} && \frac{m_D}{\sqrt{2}R} && i\frac{m_D}{\sqrt{2}R}\\
    \frac{m_DM_D\epsilon}{R^3} && \frac{1}{\sqrt{2}} && -i\frac{1}{\sqrt{2}}\\
    -\frac{m_D}{R} && \frac{M_D}{\sqrt{2} R} && i\frac{M_D}{\sqrt{2} R}
   \end{pmatrix}.
\eeq
In the $m_D\ll M_D$ limit, the active neutrino states almost 
coincide with the flavor states $\nu_i\simeq\nu_l$, and the two singlets combine into heavy physical states $N_{\pm}\simeq(N_S\pm N_R)/\sqrt{2}$. The mixing between active-hidden 
and hidden-hidden neutrino states can also be measured by a set of three characteristic angles,
\beq
 \begin{split}
  \theta_{\nu S}\approx\theta_{\nu N}\approx \frac{m_D}{\sqrt{2}M_D},\quad\theta_{NS}\approx-\frac{\pi}{4}+\frac{m_D^2}{4M_D^2}.
 \end{split}
\eeq
Because the $\nu-S$ and $\nu-N$ mixing angles are so similar, we define the angle $\Theta_i\equiv\theta_{\nu_iN}$ to characterize the visible-hidden mixing.

\begin{figure}[t!]
\includegraphics[width=8.5cm]{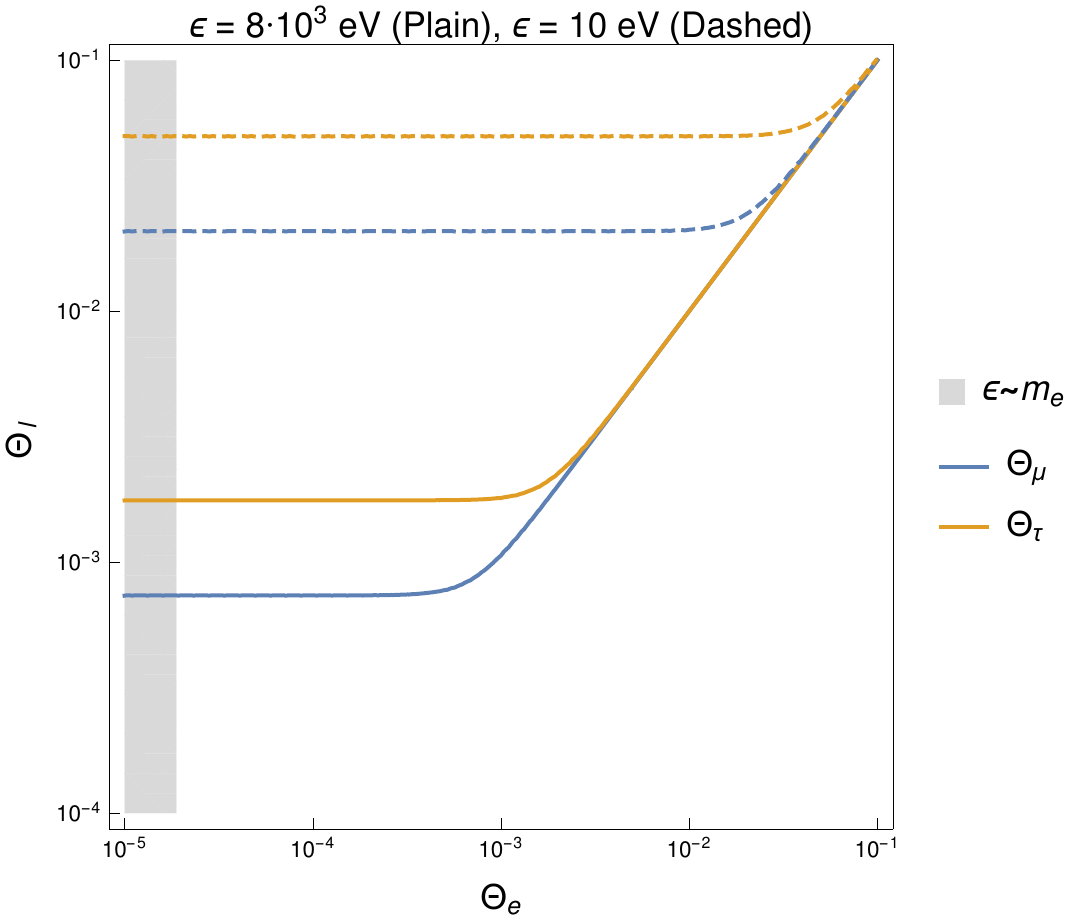}
\caption{\footnotesize A plot of the angles $\Theta_{\mu,\tau}$ given by equations \eqref{eq: theta mu and tau}, with $M_D=3$ GeV. The gray band is inconsistent with the 
assumption that $\epsilon\ll m_D$.}
\label{fig: Neutrino mass squared}
\end{figure}

We will generally make the assumption that the matrices $M_D, \epsilon$ are diagonal and universal, so that they are characterized by one scale also denoted $M_D,\epsilon$ 
respectively (this may involve some tuning, since in general it is not possible to simultaneously diagonalize both matrices). The visible Dirac masses $m_D$ are nearly diagonal, 
but not universal, so that $m_D$ can be characterized by three parameters $m_{De,\mu,\tau}$, thus $m_{\nu_l}\approx\epsilon m_{Dl}^2/M_D^2$. The first two constraints on the 
model are those from the neutrino mass squared differences, $\Delta m_{21}^2=m_{\nu_2}^2-m_{\nu_1}^2$ and $\Delta m_{32}^2=m_{\nu_3}^2-m_{\nu_2}^2$,
\beq\label{eq: theta mu and tau}
 \Theta_{\mu}^4\simeq\frac{\Delta m_{21}^2}{4\epsilon^2}+\Theta_{e}^4,\quad \Theta_{\tau}^4\simeq\frac{\Delta 
m_{32}^2+\Delta m_{21}^2}{4\epsilon^2}+\Theta_{e}^4.
\eeq
These functions have been plotted for clarity in Fig.~\ref{fig: Neutrino mass squared}. As we will see in the rest of this section, constraints can be imposed on $\Theta_e$, 
cutting off some 
of the available parameter space. Further constraints could also be placed on $\Theta_{\mu,\tau}$, 
though we will not study these explicitly in this paper (see e.g. \cite{Deppisch:2015qwa}).

\paragraph{Lepton flavor violation}

\begin{figure}[t!]
\includegraphics[width=7cm]{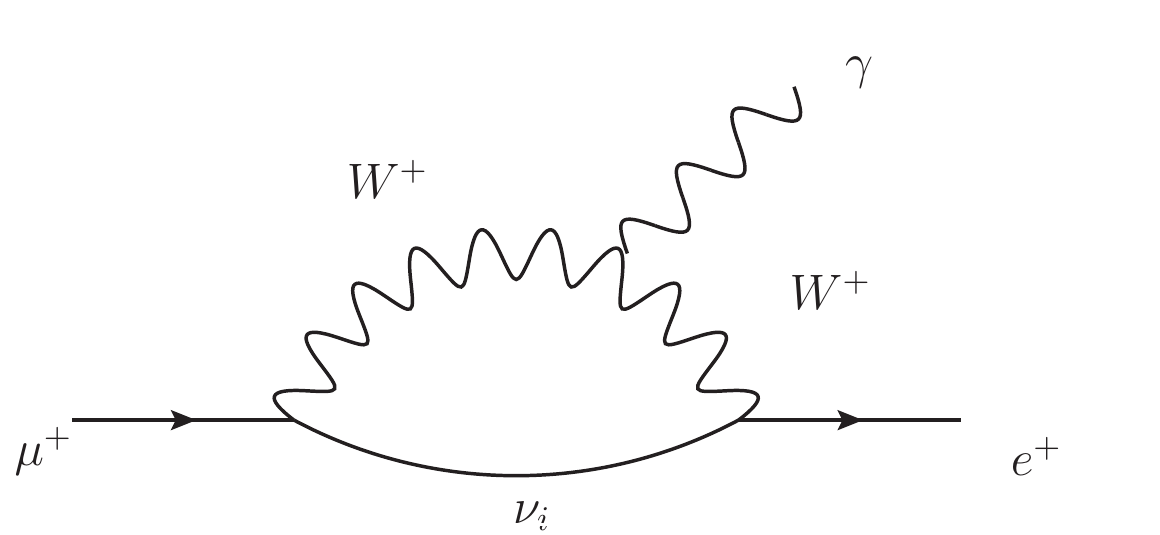}
\caption{Example of a $\mu\rightarrow e\gamma$ diagram.} 
\label{mu_to_e_gamma}
\end{figure}

One NP-sensitive lepton flavor violating (LFV) observable is the $\mu\rightarrow e\gamma$ decay, whose branching ratio is experimentally 
constrained to satisfy $\text{Br}(\mu^-\rightarrow e^-\gamma)<5.7\cdot10^{-13}$ \cite{Brooks:1999pu, Adam:2013mnn}. The MEG-II upgrade is expected to have an order of magnitude 
better sensitivity \cite{Renga:2014xra}. Within the context of the SM, this process proceeds through a $W-\nu$ loop, an example of which is shown in 
Fig.~\ref{mu_to_e_gamma}. Including corrections from the massive neutrinos, the result is 
$\text{Br}(\mu\rightarrow e\gamma)=(3\alpha/32\pi)\delta_\nu^2$ \cite{Cheng:1976uq, Bjorken:1977br, Ma:1980gm, Casas:2001sr, Ilakovac:1994kj}, with
\beq
 \delta_\nu=2\sum_{i=\nu,\pm}U_{ei}^*U_{\mu i}g\left(\frac{m_i^2}{m_W^2}\right),
\eeq
where
\beq
 g(x)=\int_0^1d\alpha\frac{1-\alpha}{1-\alpha+\alpha x}\left[2(1-\alpha)(2-\alpha)+\alpha(1+\alpha)x\right].
\eeq
In the small $x$ limit, $g(x\ll1)\approx5/3-x/2$, and using unitarity we find $\delta_\nu\approx-\sum_{i=\nu,\pm}U_{ei}^*U_{\mu i}m_i^2/m_W^2$. Furthermore, unitarity also implies 
$U_{e\nu}^*U_{\mu \nu}=-\sum_{i=\pm}U_{ei}^*U_{\mu i}$, resulting in a GIM-like dependence of $\delta_\nu$ on the mass differences 
$m^2_\pm-m^2_\nu$ \cite{Grinstein:2015nya}. Given the tiny active neutrino masses, the parameter $\delta_\nu$ is dominated by the $M_\pm$ 
contributions 
$\delta_\nu\approx-\Theta_{e}\Theta_{\mu}(M_+^2+M_-^2)/m_W^2$. Thus we have
\beq\label{eq: lepton flavor violation}
 \text{Br}(\mu\rightarrow e\gamma)\approx\frac{3\alpha}{8\pi}\frac{M_D^4}{m_W^4}\Theta^2_{e}\Theta^2_{\mu}<5.7\cdot10^{-13}.
\eeq
Inspection of this formula shows that there is ample room to saturate this inequality with light NP: taking $\Theta^2_{e}\Theta^2_{\mu} \sim 10^{-8}$  and $M_D < m_W$ allows for 
$\text{Br}(\mu\rightarrow e\gamma)$ at a level close to its upper bound. 

For this particular light NP model, stronger sensitivity actually comes from $\mu-e$ conversion in nuclei \cite{Vergados:1985pq, Kosmas:1993ch, Weinberg:1959zz}. There are two 
transition channels for this process. First, the \textit{photonic} transition, which proceeds via the same diagram as the $\mu\rightarrow e\gamma$ on-shell transition, except the 
photon is now connected to the nucleus. The 
second, the \textit{non-photonic} transition, occurs either through $Z$-boson exchange or a box diagram mediated by $W$-bosons. Within the model considered here,  the non-photonic 
transition is dominated by the box diagram. Since we focus on the light neutrino regime, we utilize the result that the box diagram receives a large enhancement (in this case) 
compared to the photonic transition, $\Gamma_{\cancel\gamma}/\Gamma_{\gamma}\sim10^3$ \cite{Vergados:1985pq}, and the box diagram dominates the $\mu-e$ conversion rate. The 
conversion rate compared to muon capture in the nucleus, $R_{\mu-e}=\Gamma(\mu-e)/\Gamma_{capture}$, is given by
\beq
 R_{\mu-e}\approx\left(\frac{3G_Fm_W^2}{4\sqrt{2}\pi^2}\right)^2\frac{E_ep_e}{m_\mu^2}|F_{ch}|^2\rho\times\delta_\nu^2,
\eeq
where
\beq
 \rho\approx Z\frac{|3/2\beta_0(1+N/Z)+\beta_1/2(1-N/Z)|^2}{6|1.62Z/A-0.62|},
\eeq
is an enhancement factor accounting for the coherent nature of the transition. The charge form factor can be experimentally determined for various elements, the largest of 
which is about $|F_{ch}|\approx0.5$ \cite{Donnelly:1975ze, Weinberg:1959zz}. The parameters $\beta_0\sim30$, $\beta_1\sim25$, and the factor $E_ep_e/m_\mu^2\approx1$. The best 
current limit is $R_{\mu-e}\Leq7.0\cdot10^{-13}$ from experiments using gold $^{197}\text{Au}$ \cite{Bertl:2006up}, for which $Z=79$, and the coherent enhancement factor is 
$\rho\approx1.6\cdot10^{6}$. Therefore, using the above expression 
for $\delta_\nu$, we have
\beq
 R_{\mu-e}\approx6.5\delta_{\nu}^2\approx26\frac{M_D^4}{m_W^4}\Theta^2_{e}\Theta^2_{\mu}\Leq7.0\cdot10^{-13},
\eeq
which is a stronger constraint on mass and mixing by two orders of magnitude than the $\mu \rightarrow e\gamma$ branching ratio. Various experiments, either running or in the 
planning stages, aim
to increase sensitivity by several orders of magnitude \cite{Natori:2014yba, Kutschke:2011ux}.
This inequality can again be saturated with $M_D \ll m_W$.

\begin{figure}[t!]
\includegraphics[width=8cm]{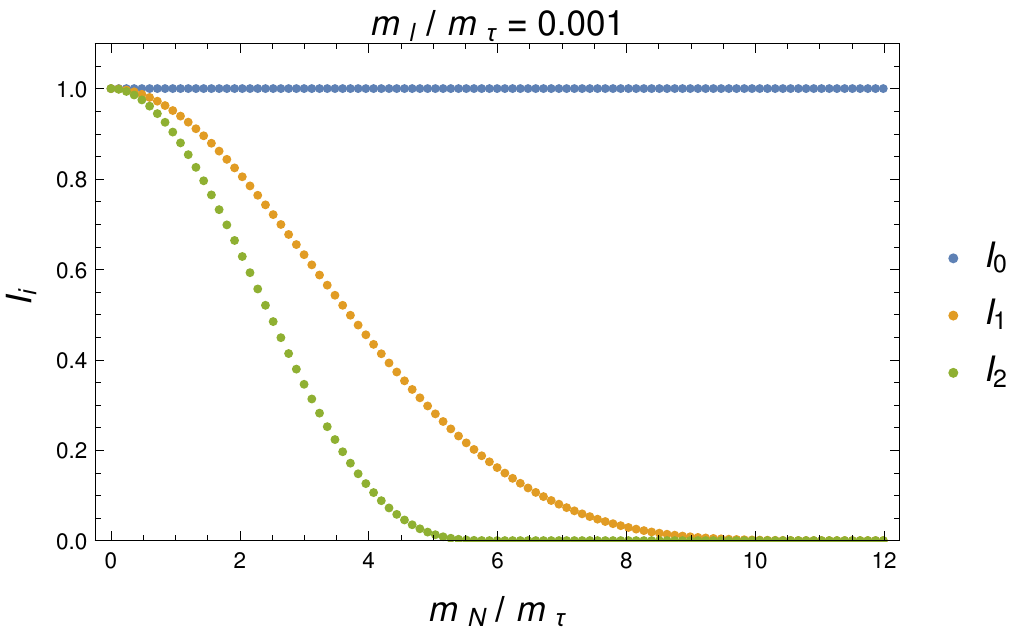}
\caption{\footnotesize Phase space factors $I_{0,1,2}$ for the $\tau\rightarrow l\nu\nu$ decay.} 
\label{fig: phase space integrals}
\end{figure}

\paragraph{Lepton Universality}
Tests of lepton universality provide valuable constraints on masses and mixings of massive neutrinos. 
There exist various standard decay channels to test lepton universality (see {\em e.g.} \cite{Lusiani:2007cb, Pich:1997hj,Ilakovac:1994kj}). 
In particular, we will focus on $\mu-e$ universality in $\tau$ decays, through 
the $R_\tau$ observable defined as
\beq
 R_\tau=\frac{\Gamma(\tau^-\rightarrow e^-\nu\nu)}{\Gamma(\tau^-\rightarrow \mu^-\nu\nu)}.
\eeq
In the SM, because the neutrinos are massless, the flavor eigenstates $\nu_l$ and mass eigenstates $\nu_i$ coincide so that $\Gamma(\tau^-\rightarrow 
e^-\nu\nu)=\Gamma(\tau^-\rightarrow e^-\nu_\tau\nu_l)$. For massive neutrinos, and multiple neutrino states, the masses are linear combinations of flavor eigenstates, 
and $\Gamma(\tau^-\rightarrow 
e^-\nu\nu)=\sum_{i,j}\Gamma(\tau^-\rightarrow l^-\nu_i\nu_j)$. The $R_\tau$ ratio has recently been measured by BaBar, 
$R_\tau=0.9796\pm0.0016\pm0.0036$ \cite{Aubert:2009qj}, which we will approximate as $R_\tau\approx1\pm\Delta R_\tau$, with $\Delta R_\tau=0.0052$. 
In general, the decay rate takes the form
%have a somewhat cumbersome dependence on the various lepton and neutrino masses, depending on how many final states are kinematically accessible in a given decay channel 
\cite{Dib:2011hc, Abada:2013aba},
\beq
 \Gamma(\tau^-{\rightarrow}l^-\nu\nu)=\frac{G_F^2m_\tau^5}{192\pi^2}\sum_{ij}|U_{\tau i}|^2|U_{lj}|^2I\left(\frac{m^2_l}{m^2_\tau},\frac{m^2_{\nu_{i,j}}}{m^2_\tau}\right).
\eeq
To a good approximation, we 
can take the active neutrino masses to vanish, and the splitting between the two sterile neutrino states to be negligible, $M_+-M_-\sim\epsilon\ll m_\tau$. Thus, the 
kinematic function $I$ splits into three categories, where either zero, one or two of the sterile states are produced, respectively denoted $I_{0,1,2}$, with each 
depending only on the mass scales, but not the flavors. These are plotted in Fig.~\ref{fig: phase space integrals}, and are very insensitive to the outgoing lepton masses $m_\mu$ 
and $m_e$. As a result,
%%%%
%%%%
\begin{align}
  \Gamma(\tau^-{\rightarrow}l^-\nu\nu)&\,{\propto}\,I^l_0\sum_{i,j}|U_{\tau \nu_i}|^2|U_{l\nu_j}|^2{+}I^l_2\sum_{i,j}|U_{\tau 
N_i}|^2|U_{lN_j}|^2 \nonumber\\
&\;\;+I^l_1\sum_{i,j}\left(|U_{\tau N_i}|^2|U_{l\nu_j}|^2+|U_{\tau \nu_i}|^2|U_{lN_j}|^2\right).
\end{align}
Using unitarity, we express the visible-visible 
mixing as $\sum_j|U_{l\nu_j}|^2=1-\sum_{i}|U_{lN_i}|^2\approx1-N_h\Theta_l^2$,  using the assumption $U_{lN_i}=\Theta_l$, with $N_h$ the number of 
hidden flavors. This approximation is valid as long as $1-N_h\Theta_l^2>0$. The constraint comes from requiring that $\Delta R_\tau=1-\Gamma(\tau\rightarrow 
\mu\nu\nu)/\Gamma(\tau\rightarrow e\nu\nu)<0.0052$ be within the experimental errors. 

The actual significance of this constraint depends on the concrete realization of the mass and mixing pattern. 
For example, in the situation where the heavy neutrino eigenstates cannot be kinematically produced, the final constraint can be presented as 
$N_h|\Theta_\mu^2-\Theta_e^2|<10^{-2}$. A somewhat stronger universality constraint can be derived by comparing charged pion decay modes. 
These constraints can be saturated with $\Theta_l^2\sim O(10^{-2})$ and, as we saw above, such a mixing pattern can easily arise from new singlet neutrino 
states with a mass below the electroweak scale. 

\begin{figure}[t]
\flushleft
\subfigure[]{
\includegraphics[width=7.2cm]{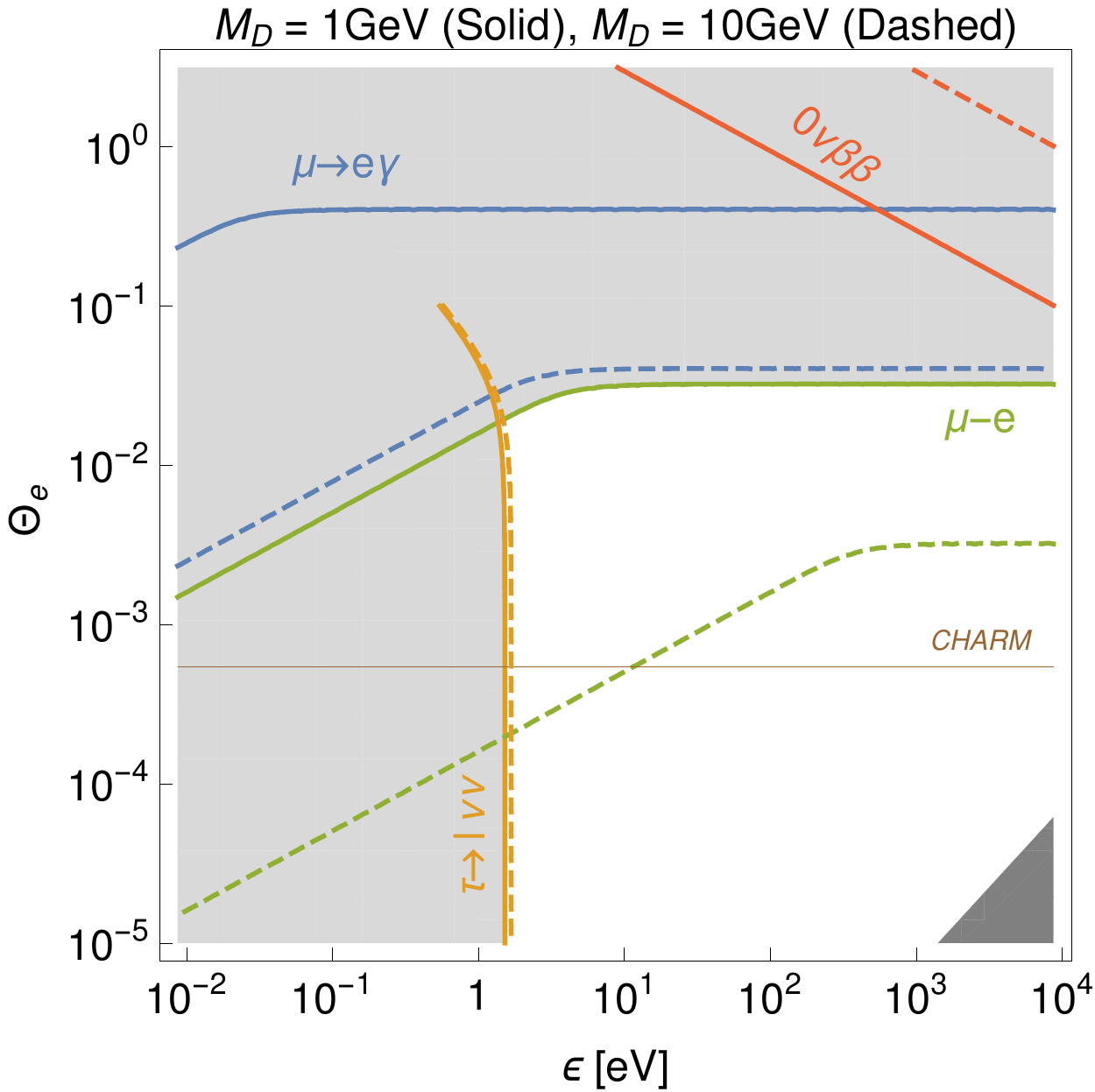}
\label{fig: Paramater space constraints low M}}
\subfigure[]{\includegraphics[width=7.2cm]{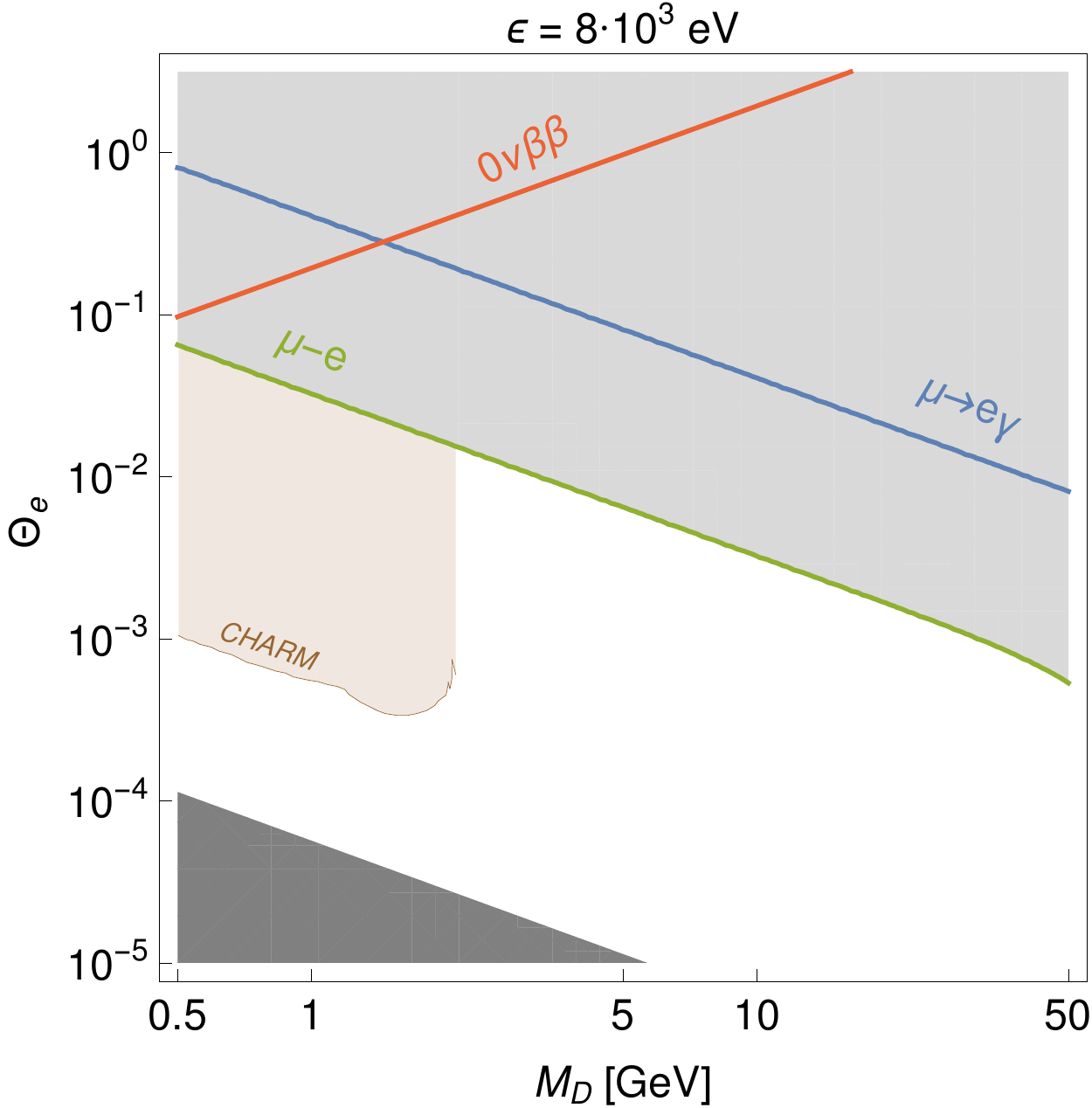}
\label{fig: Paramater space constraints high epsilon}}
\caption[]{Plots showing the allowed regions in (a) $\{\epsilon,\Theta_e\}$ and (b) $\{M_D,\Theta_e\}$ respectively, with $N_h=2$. The lines correspond to saturation of the 
respective bounds. $M_D$ is bounded above by the requirement $M_D<m_W$, and from below by the $0\nu\beta\beta$ constraint $M_D\gg 0.1$~GeV.  We note that at large $\epsilon$, the 
$\mu-e$ conversion constraint is by far the strongest, though universality becomes stronger at low $\epsilon$. The darker gray area represents the regime where 
$\epsilon/m_e>10^{-1}$, with $\epsilon$ no longer `small'. For completeness, we also show a constraint from on-shell production of sterile states at the CHARM fixed target 
experiment \cite{CHARM}, which applies for $M_D \lsim 2\,$GeV.}
\label{fig: Paramater space constraints}
\end{figure}

In summary, using the neutrino mass differences to express $\Theta_{\mu,\tau}$ as functions of $\epsilon,\Theta_e$, we can present the LFV and universality constraints above 
in the parameter plots shown in Fig.~\ref{fig: Paramater space constraints}. It is clear that this simple light NP model, with sub-EW scale singlet 
fermionic states, can induce deviations in $\mu \rightarrow e$ conversion or lepton universality at the level of the current experimental sensitivity.
Therefore, if future experiments detect a non-zero result, further work will be required to unambiguously differentiate between light and heavy NP models.

\subsection{Lepton number violation}

Neutrinoless double beta decay $\bbnu$ is the primary observable for lepton number ($L$) violation, and the current limits 
\cite{KlapdorKleingrothaus:2006ff,Bakalyarov:2003jk,Gando:2012zm, Auger:2012ar, Garfagnini:2014nla, Palioselitis:2015nna} are normally interpreted directly in terms of the light 
neutrino mass spectrum 
\cite{Schwingenheuer:2012zs,Dev:2013vxa}. Given the existing mass limits on the light eigenstates, the decay rate depends on the effective Majorana mass $m_{\nu}^{\rm eff} = 
\sum_i 
U_{ei}^2 m_i$. Even without performing a dedicated analysis, it is clear that any future evidence for a non-zero $m_{\nu}^{\rm eff} $ will not be able to differentiate between 
light and heavy NP models. Indeed, in the simplest Type I see-saw model, $m_{\nu}^{\rm eff} \sim {\cal O}({\rm eV})$ can arise from models with {\em e.g.} $M_R \sim 1$ GeV or $M_R 
\sim 10^{10}$ GeV, and therefore both interpretations would be possible.

In what follows, for completeness, we analyze lepton number violation utilizing the same neutrino model as in the previous section. Recall that the heavy mass eigenstates $N_i$ mix 
with $\nu_e$, so that the mass eigenstates 
can be written as $N_i = U_{ei} \nu_e +\cdots$. It will be sufficient to work with the following analytic approximation for the decay rate \cite{Bamert:1994qh},
\be 
 \Gamma(\bbnu) \sim \frac{G_F^4Q^5 \cos^4\theta_C}{60\pi^3}|{\cal M}|^2,
\ee
where $Q+M(Z,A)-M(Z+1,A)$ is the endpoint energy, and the amplitude takes the approximate form,
\begin{align}
 {\cal M} &= \sum_i U_{ei}^2 m_i \int \frac{d^4p}{(2\pi)^4} \left( \frac{w(p_0,|\vec{p}|)}{p^2 -m_i^2+i\ep}\right) \nonumber\\
   & \longrightarrow \; \frac{iE_Fp_Fw_0}{4\pi^3} \left\{ \begin{array}{cc} m_{\nu}^{\rm eff} + \cdots & {\rm for} \; p_F \gg m_i,\\
                               \frac{p_F^2}{3}\sum_i \frac{U_{ei}^2}{m_i} + \cdots & {\rm for} \; p_F \ll m_i. \end{array} \right.
\end{align}
The nuclear form factor $w(p_0,|\vec{p}|)$ has been approximated by a step function $w\sim w_0 \Theta(p_0 - E_F)\Theta(|\vec{p}|-p_F)$, with $w_0 \sim 4$~MeV$^{-1}$  
\cite{Bamert:1994qh} in terms of the nucleon Fermi momentum $p_F\sim 100$~MeV. (A more precise interpolating formula is given in \cite{Faessler:2014kka}.)

When the dominant contribution is from the light active neutrinos, the experimental bounds translate to $m_{\nu}^{\rm eff}\lesssim0.12-0.38\rm eV$ \cite{Gando:2012zm, 
Auger:2012ar}. 
It is also instructive to separately estimate the sensitivity to the exchange of heavier neutrino eigenstates. 
Note that when the singlet mass $m_N \gg p_F$, 
the experimental constraints lead to a bound on $(p_F^2/3)\sum_iU_{ei}^2/m_i$ instead. Within the neutrino mass model described in the previous section, we find
\beq
 \sum_i \frac{U_{ei}^2}{m_i}\simeq\frac{U_{e+}^2}{M_+}+\frac{U_{e-}^2}{M_-}\simeq\Theta_e^2\left(\frac{M_--M_+}{M_+M_-}\right),
\eeq
where $M_--M_+\simeq-\epsilon$, and $M_-M_+\simeq M_D^2$. Thus, in the regime $M_D\simeq 1~\rm GeV$, we use the bound 
$0.3(p_F/M_D)^2\Theta_e^2\epsilon\lesssim0.3\rm eV$. 

The above bound is displayed in Fig.~\ref{fig: Paramater space constraints} for comparison with the LFV constraints. LNV provides a subleading constraint within this particular 
inverse seesaw model, since the lepton number violating parameter is $\epsilon$ which is taken to be small compared to the other mass scales. However, as already emphasized above, 
more significant sensitivity to the mixing angle arises in the standard seesaw model, where we enlarge the Majorana terms in the mass matrix.

\subsection{Electric dipole moments}

Electric dipole moments (EDMs) constitute an important class of precision $CP$-odd observables. 
There are several channels by which $CP$ violation can be communicated from the light NP degrees of freedom to the 
SM within a UV-complete model. $CP$-odd mediation can occur via the neutrino portal, and the phases in the Yukawa matrix $Y_N$, provided there are at least two singlet neutrinos 
$N_i$. The second channel is the Higgs portal. While obviously $CP$-even by itself, the scalar mediator can couple
to a light NP sector in a manner that explicitly breaks $CP$, and this may be communicated to the SM via higher-order loop effects. 
We will first consider lepton EDMs induced through the neutrino portal before turning to more generic light NP mechanisms, and subsequently consider hadronic EDMs, which are 
distinct in that they can be generated at or close to the current level of sensitivity through the QCD $\theta$-term.

\paragraph{Paramagnetic (and leptonic) EDMs}

\begin{figure}[t!]
\includegraphics[width=8cm]{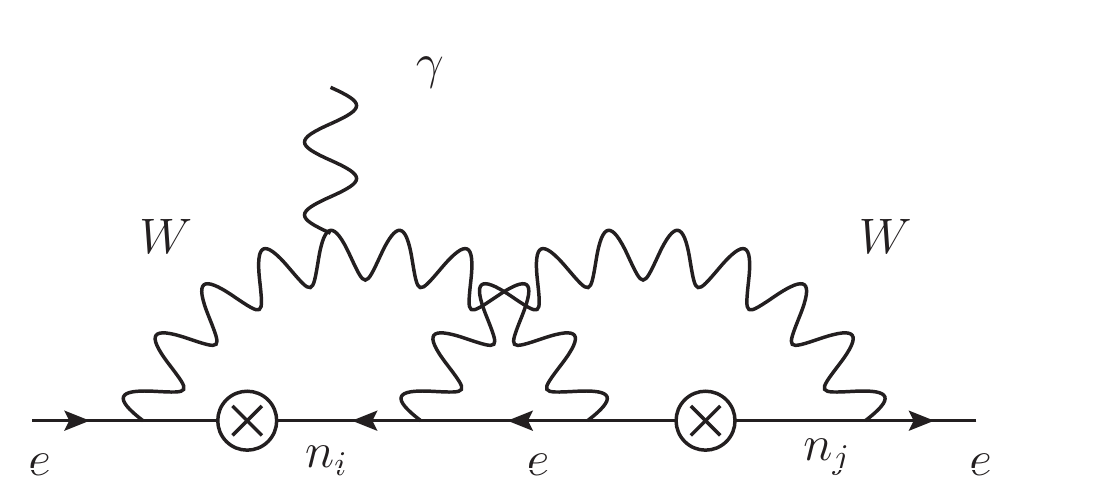}
\caption{An example of a W-loop diagram contributing to the lepton EDM. The crosses indicate neutrino mass insertions.} 
\label{fig: neutrino edm}
\end{figure}

We first consider the same neutrino mass matrix studied in the previous subsection.
In the SM extended with massive neutrinos, it is possible to generate an EDM at the two-loop level \cite{Ng:1995cs, Archambault:2004td, deGouvea:2005jj}, an 
example of which is shown in Fig.~\ref{fig: neutrino edm}. These diagrams can be shown to scale as
\beq
\begin{split}
 d_e\sim 
&~em_e\left(\frac{G_F}{16\pi^2}\right)^2\\&\times\sum_{i<j}\Gamma_{ij}m_im_j\frac{\left(m_j^2-m_i^2\right)}{m_W^2}F\left(\frac{m_i}{m_W},\frac{m_j}{m_W},\frac{m_e}{m_W} 
\right) ,
\end{split}
\eeq
where $\Gamma_{ij}=\text{Im}\left\{(U^*_{ei})^2U^2_{ej}\right\}$, and $i,j=\nu,\pm$ are the neutrino mass eigenstates, and $F$ is a loop function. Up to now, we have ignored the  
possible $CP$-odd phases in this model. However,  in general not all the mass parameters are real. We choose to leave the physical $CP$-odd phase in $m_D=|m_D|e^{i\eta}$, and to 
ease the notation, replace $|m_D|$ simply with $m_D$. Given that the mixing angles are $m_D/M_D\lesssim0.1$, we see that the 
contribution from the diagram with two internal light neutrinos will be tiny, $~\mathcal{O}(m_{\nu_1}m_{\nu_2}\Delta m_{21}^2)$. Thus, we need only look at the cases where either  
one or two internal neutrinos are heavy, respectively called the $h-l$ or $h-h$ contributions. Looking at the $h-h$ contribution, the mixing that enters is $\Gamma_{+-}$, but both 
$U^2_{e+}$ and $U^2_{e-}$ have the same phase since they are controlled by $m_D/M_D$. The $CP$-odd phase thus cancels from the $h-h$ contribution. Next,  the $h-l$ contribution is 
proportional to $\Gamma_{\nu_e\pm}\sim\pm m_D^2/(2M_D^2)\sin(2\eta)$, whereas $m_{\nu_e}M_\pm(M_\pm^2-m_{\nu_e}^2)\sim m_D^2M_D\epsilon\pm3/2m_D^2\epsilon^2$. As a result, the 
$h-l$ contribution to the EDM vanishes at $\mathcal{O}(\epsilon)$. So at the lowest non-vanishing order, we have
\beq
 d_e=d_e^{h-l}\sim \Theta_e^4\frac{M_D^2}{m_W^2}\frac{\epsilon^2}{\Delta m^2_{21}}\cdot 10^{-53}\sin(2\eta)e\, \text{cm},
\eeq
assuming the function $F\left(\frac{m_{\nu_e}}{m_W},\frac{M_\pm}{m_W},\frac{m_e}{m_W}\right)$ is of order unity. Within the allowed parameter space of  Fig.~\ref{fig: Paramater 
space constraints}, the above EDM is maximal for  $\epsilon^2/\Delta m^2_{21}\lesssim10^{12}$, $M_D/m_W\sim10^{-2}$ and $\Theta_e\lesssim10^{-2}$, leading to an upper bound 
$d_e<10^{-53}e\cdot \text{cm}$. The suppression of the upper bound arises from the 
size of the Majorana mass term $\epsilon$, which is set by the constraint on the active neutrino mass squared differences. Therefore, within this model it is not possible to  
generate a sizeable EDM. However, a far larger EDM is possible in a variant of this model with an extra visible-hidden Dirac mass coupling $m_2$. Namely, we switch gear and 
consider the following extended mass matrix,
\beq
 -\mathcal{L}_{\nu}\supset\left(\nu_L\quad N_R\quad N_S\right)\begin{pmatrix}
                    0 && m_{D_1} && m_{D_2}\\
                    m_{D_1} && M_R && \epsilon\\
                    m_{D_2} && \epsilon && M_S
                   \end{pmatrix}\begin{pmatrix}
                                 \nu_L\\ N_R\\ N_S
                                \end{pmatrix},
\eeq
in the regime $M_{R,S}\gg m_{D_i},\epsilon$. The limiting case $\epsilon=M_R=0$ leads to two light neutrinos, 
and only one heavy neutrino, and will not lead to enhanced EDMs. Therefore we are forced to consider the full spectrum, and treating $\epsilon$ 
as a perturbation leads to
\beq
\begin{split}
 &m_\nu\approx m_\nu^0+2\frac{m_{D_1}m_{D_2}}{M_RM_S}\epsilon,\\
 &M_\pm\approx M_\pm^0\pm\frac{\epsilon}{\Delta M}\left(\epsilon+2\frac{m_{D_1}m_{D_2}}{M_\pm}\right),
\end{split}
\eeq
where $m_\nu^0\simeq(m_{D_1}^2-m_{D_2}^2)/M$, $M_+^0\simeq M_S$, $M_-^0\simeq M_R$, $\Delta M=M_S-M_R$ is the Majorana mass splitting, and 
$M=(M_R+M_S)/2$ is the Majorana mass scale.  Even though we are in a see-saw-like 
scenario with large Majorana masses, $M_D$ easily evades the light neutrino mass constraints since they are now controlled by the fine tuning of $m_{D_2}^2-m_{D_1}^2$. For 
simplicity we have ignored the phase in the mass eigenstates. As a consequence, to lowest order in $\epsilon(\sim0)$ \cite{Archambault:2004td},
\beq
\begin{split}
 d_e\sim 
&~ em_e\left(\frac{G_F}{16\pi^2}\right)^2\frac{\Delta M}{M}\frac{m_{D_1}^2m_{D_2}^2}{M^4}M^2\\&\times\left(\frac{32}{3}
\ln\left(\frac {M}{M_W} \right)-\frac { 260 } { 9 } +\frac{112}{27}\pi^2\right)\sin(2\eta),\\
\sim&\left(3\cdot10^{-35}~ e\,\text{cm}\right)\frac{m_{D_1}^2m_{D_2}^2}{M^4}\frac{M_S^2-M_R^2}{\text{GeV}^2}\sin(2\eta).
\end{split}
\eeq
The ratios $m_{D_i}/M\lesssim10^{-1}$ are the visible-hidden mixing angles. Thus, on choosing a 
mass scale $M_S^2-M_R^2\sim M_S^2\simeq10^2~\text{GeV}^2$, one finds $d_e\lesssim10^{-37}e\cdot \text{cm}$. Allowing for significant fine tuning, it is possible to enhance this 
upper bound to $\sim10^{-33}\text{e}\cdot\text{cm}$, which is still considerably lower than the current experimental upper limit
$d_e < 8.7\times 10^{-29}e\cdot \text{cm}$ \cite{Baron:2013eja}. Therefore, we conclude that the sterile neutrino $CP$ violating portal falls short of inducing $d_e$  close to the 
current experimental bound.

Taking a more general approach, we now consider a more complex light hidden sector. We introduce a Dirac fermion $\ps$ charged under $U(1)_V$ with $CP$-violating couplings to the 
scalar singlet $S$ \cite{Bird:2006jd,O'Connell:2006wi,Barger:2007im,Sato:2011gp,Fox:2011qc,Low:2011gn},
\begin{align}
{\cal L}_{\rm hid}^{\rm CP} = \bar\ps i\gamma^\mu D^V_\mu\psi + \bar\psi (m_\psi + S(Y_S + i\tilde{Y}_S \gamma_5))\ps.
\end{align}
where $D^V_\mu = \ptl_\mu - e'q_\psi V_\mu$. This hidden sector $CP$-violation can then be mediated to the SM via the $CP$-even vector and Higgs portals.

\begin{figure}
\includegraphics[viewport=160 450 440 720, clip=true, scale=0.4]{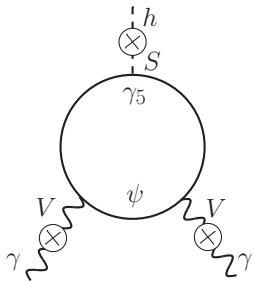}
\caption{The $CP$-odd $hF\tilde{F}$ vertex. } 
\label{fig1}
\end{figure}

Loops of $\psi$ can  induce the $CP$-odd  $SV\tilde{V}$ vertex shown in Fig.~\ref{fig1}, which will contribute to EDMs at higher loop order, via e.g. Barr-Zee-type diagrams. For 
example, integrating out the scalar and one of the vector legs, one obtains a `dark $V$-EDM' operator of the electron, which will in turn translate to the effective `EDM radius' 
operator (analogous to the charge radius). Schematically, this process of integrating out short distance scales can be presented as
\be
S V_{\mu\nu} \tilde V^{\mu\nu}\to \frac{i}{2}\bar \psi_e \sigma^{\mu\nu} \gamma_5 \psi_e V_{\mu\nu} \to \frac{i}{2} \bar \psi_e \sigma^{\mu\nu} \gamma_5 \psi_e \frac{\Box 
F_{\mu\nu}}{m_V^2}.
\ee
Denoting the coefficient in front of $\frac{i}{2} \bar \psi_e \sigma^{\mu\nu} \gamma_5 \psi_e\Box F_{\mu\nu}$ as 
$r^2_{ d_f}$, we can utilize existing EDM calculations. For simplicity, we assume 
that $V$ is parametrically lighter than $S$ and $\ps$, so the EDM radius takes the form (see e.g. \cite{McKeen:2012av})
\be
\label{2loopSh}
 r^2_{d_f} = \frac{|e| \alpha' \tilde{Y}_S m_f}{16\pi^3 v m_\ps m_V^2} \ka^2 \sin(2\theta)\left[ g(m_\ps^2/m_h^2)-g(m_\ps^2/m_S^2) \right],
\ee
where the loop function is given by
\be 
 g(z) = \frac{z}{2} \int_0^1 dx \frac{1}{x(1-x)-z} \ln\left(\frac{x(1-x)}{z}\right),
 \label{eq:g}
\ee
and satisfies $g(1)\sim1.17$, $g(z\ll 1) \sim z(\ln z)^2/2$ and $g(z\gg1) \sim \frac{1}{2}\ln z$. Within the fully hierarchical regime,
\be
m_V \ll m_S \ll m_\ps,
\ee
and taking $\theta \sim -Av/(m_h^2) \ll 1$, we have
\be
   r^2_{d_f}  = - \frac{|e| \alpha' \tilde{Y}_S m_f}{16\pi^3 v m_\ps m_V^2} \times  \ka^2 \theta  \ln(m_\ps^2/m_S^2).
\ee
This operator leads to the usual $s-p$ mixing of atomic orbitals, and the atomic EDM linked to the electron spin direction. 
While a full atomic calculation is required to deduce the size of the matrix element, we will resort to estimating its size 
by saturating $\Box$ with the square of the inverse radius of a K-shell.  Then, the effective EDM radius translates to an electron EDM of size $d_e^{\rm eq} \sim (Z \alpha m_e)^2  
r^2_{d_e}$, as long as $m_V > Z \alpha m_e$. 
Taking $m_V \sim (m_e \al Z)$, 
$q_\ps=1$, $\al'=\al$, $\tilde{Y}_S=1$ and taking the log to be ${\cal O}(1)$, we arrive at the estimate
\be
 d_e^{\rm eq}  \sim 4\times 10^{-33}~e\cdot {\rm cm} \times \left(\frac{1\,{\rm GeV}}{m_\psi}\right)\left(\frac{\ka}{10^{-4}}\right)^2 \left( \frac{\theta}{10^{-3}}\right),
\ee
which is still well below the current sensitivity to the electron EDM \cite{Baron:2013eja}. The main difficulty is in taking $m_V$ in the sub-MeV range, 
where $g-2$ of the electron imposes a strong limit on kinetic mixing, $\ka < 10^{-4}$. We note in passing that extracting a proper limit on the `dark $V$-EDM', or $ r^2_{d_e}$, is 
a well-motivated problem for atomic physics, in line with the recent investigation of $CP$-odd operators induced by a mediator of mass comparable to the inverse atomic scale 
\cite{Gharibnejad:2014kda}.

Another contribution to the experimentally accessible paramagnetic EDMs of atoms and molecules is the semi-leptonic interaction $C_S \bar{N}N \bar{e} i \gamma_5 e$. This operator 
can also be generated through the $hV\tilde{V}$ vertex, and allows access to a regime with larger $m_V$ for which the constraints on $\ka$ are somewhat weaker. However, this 
contribution is still not at a level that can approach the current experimental sensitivity. Thus, at least within this restricted class of hidden sectors, we conclude that 
paramagnetic EDMs (and specifically lepton EDMs) are in practice a probe of UV new physics.

\paragraph{Diamagnetic (and hadronic) EDMs}

Hadronic EDMs can also be induced using the mechanism outlined above, via the hidden sector Barr-Zee diagram, although again necessarily  below the current level of sensitivity. 
However, hadronic EDMs can also be generated by the QCD $\theta$-term, which is a marginal operator 
\cite{Baluni:1978rf,Crewther:1979pi,Pospelov:2000bw,Pospelov:1999ha}. The strongest current limits in this sector are from the EDM of the neutron \cite{Baker:2006ts}
\be
|d_n| < 2.9 \times 10^{-26} e\,{\rm cm},
\ee
and the EDM of the Hg atom \cite{Griffith:2009zz},
\be
 |d_{Hg}| < 3.1 \times 10^{-29} e\,{\rm cm},
\ee
where the apparent strength of the Hg EDM bound is tempered by Schiff screening of the nuclear EDM. The contribution of $\bar\theta$ to the neutron EDM takes the form (see e.g. 
\cite{PR}),
\be
 d_n(\bar\theta) \sim 3\times10^{-26} \left(\frac{\bar\theta}{10^{-10}}\right)\, e{\rm cm}.
\ee
This leads to the current constraint of $\bar\theta< 10^{-10}$. The contribution of $\bar\theta$ to $d_{Hg}$ is more complex, and for some time it appeared that it would be 
isospin-suppressed, with the Schiff moment for Hg primarily sensitive to the $CP$-odd isovector pion-nucleon coupling $\bar{g}^1(\bar\theta) \sim 0.001 \bar\theta$ rather than the 
isoscalar coupling $\bar{g}^0(\bar\theta)\sim 0.05 \bar\theta$. However, more recent analyses of the Hg Schiff moment have indicated that $\bar{g}^0$ may provide a comparable 
contribution to $\bar{g}^1$ \cite{Ban2010}. Taking the current `best values' \cite{Engel:2013lsa} indicates that 
\be
 d_{Hg}(\bar\theta) \sim 5\times 10^{-30} \left(\frac{\bar{g}^0(\bar\theta)}{0.05 \times10^{-10}}+{\cal O}(\bar{g}^1(\bar\theta))\right)\, e{\rm cm}.
\ee
This is a factor of 6 below the current bound, but given that the precision of the calculation is generally understood to be at the order of magnitude level 
\cite{PR,Engel:2013lsa}, it is clear that a  nonzero detection of $d_{Hg}$ could not unambiguously be distinguished from the effect of nonzero $\bar\theta$. 

We conclude that, at current levels of sensitivity, nonzero detections of the dominant hadronic EDM observables could be explained without additional UV new physics, simply through 
$CP$-odd QCD physics in the form of $\bar\theta$. Further improvements in the sensitivity to $d_n$ could of course change this picture.

\subsection{Hadronic flavor violation}

We now turn to precision quark flavor-violating observables, with the $b\rightarrow s\gamma$ transition as a benchmark. There are a couple of features which clearly distinguish 
these observables, particularly concerning the role of light new physics. Firstly, since the RH states are charged, there is no analogue of the neutrino portal, and thus no 
renormalizable flavor-violating interactions that do not involve new charged states. Given the existing limits on new light degrees of freedom which are charged, this pushes 
hadronic flavor violating observables into a category that is primarily sensitive to UV new physics. Having said this, the second distinguishing feature is that the SM itself 
provides non-negligible contributions to hadronic flavor violation through the CKM matrix. This allows for new flavor violating transitions to occur on introducing purely 
flavor-diagonal light NP. An example is the kinetic mixing between $V$ and $\gamma$ (or $Z$) in (\ref{portal}) that can induce flavor-violating transitions of the form 
$b\rightarrow sV$ and thus $b\rightarrow s\gamma^*(Z^*)$. However, given the constraints on kinetic mixing, this mechanism is too weak to produce sizeable effects without 
additional $V-Z$ mass mixing that in turn requires new UV physics \cite{DLM}. Moreover, utilizing a $W$-boson loop at leading order for example, the SM contributes to 
$\text{Br}({\bar B}\rightarrow X_s\gamma)=(3.60\pm0.30)\cdot10^{-4}$ \cite{Gambino:2001ew}, while the BaBar sensitivity is $\text{Br}({\bar B}\rightarrow 
X_s\gamma)\simeq(3.15\pm0.23)\cdot10^{-4}$ \cite{Lees:2012wg}. Thus, the sensitivity to new physics is also limited by the precision of SM calculations.

Since these observables are not primarily sensitive to light new physics in the categories that we have delineated, we will not consider them in detail. However, it is worth 
outlining how a model with low energy flavor violation can be realized, albeit one that still relies on additional heavy charged states for consistency. We consider a model in 
which we gauge an anomaly-free $U(1)'$ combination of quark flavors, $Q_{f1} - Q_{f2}$ \cite{Batra:2005rh, Dobrescu:2014fca}, with a diagonal (and vectorial) gauge coupling of the 
form \cite{Nir:1990yq,Langacker:2000ju,Carena:2004xs,Salvioni:2009jp},
\be
 {\cal L}_{Z'} = g_z Z'_\mu \sum_{q=Q_f,u_f,d_f} z_f \bar{q} \gamma^\mu q,
\ee
with e.g. $z_{Q_3} = z_b=- z_{Q_2} = -z_s = 1$. On transforming to the mass eigenstate basis, this non-universal coupling will generate a flavor-violating $b-s-Z'$ vertex, and 
mediate sizable flavor violating transitions. In practice, this imposes significant constraints on the combinations of $g_z z_{Q,b,s}$, and is usually used to motivate 
flavor-universal $U(1)'$ charge assignments. Here, we are interested in having a light $Z'$ that can indeed mediate these transitions at the level to which current experiments are 
sensitive. If we assign integer charges, then a small gauge coupling $g_z \sim 10^{-5}$ will be sufficient for this purpose. As has recently been emphasized 
\cite{Dobrescu:2014fca}, $K^0- \bar{K}^0$ mixing requires $g_z |z_{Q_2}-z_{Q_1}| < 10^{-5} M_{Z'}/(1\,{\rm GeV})$, while $B^0-\bar{B}^0$ mixing imposes similar constraints on $g_z 
|z_{Q_3}-z_{Q_1}|$. An explicit $Z'$ model 
based on `horizontal' flavor symmetries, related to possible anomalies in $B^0$ decays, can 
be found in \cite{Crivellin:2015lwa}.

The gauging of flavor non-universal symmetries leads to further model building requirements for the quark mass spectrum, as the Yukawa matrices are now subject to additional 
constraints. Additional charged Higgs fields are required, which necessarily lie above the EW scale given the current LHC constraints. Thus, while quark flavor-violation could be 
mediated via a light $Z'$, the model would necessarily involve charged states above the EW scale, and thus UV new physics.

\subsection{Baryon number violation}

New non-SM sources of baryon number violation, for which the primary precision experiments are searches for proton decay, necessarily require new UV physics. The minimal baryonic 
vertex that converts two quarks into an anti-quark and a lepton corresponds to a higher-dimensional operator that {\em requires} new charged states in any UV completion of which we 
are aware. Therefore any detection of proton decay or $n-\bar n$ oscillation will most likely point to 
the weak scale or above, as a possible source of baryon number violation. An alternative means of introducing a low energy mediation mechanism is to gauge $B$ (see e.g. 
\cite{FileviezPerez:2010gw}) within a more extended gauge group, and rely on a new non-perturbative sector to break this new gauge symmetry.  
This, however, may not necessarily lead to any proton decay, or any other baryon number-violating observable without the participation of new charged fields in the UV.

\section{Conclusions}

Empirical evidence for new physics, e.g. neutrino oscillations or dark matter, does not always provide us with much guidance as to a natural mass or energy scale. As noted in the 
Introduction, there are UV and IR scenarios for neutrino oscillations, both of which are currently viable. Sometimes it is argued that the evidence for dark matter points to NP at 
a UV scale at or above the electroweak scale (i.e. via the `WIMP miracle'). While this may be true for examples such as the MSSM neutralino or the QCD axion, which require UV 
completion with states above the EW scale, there are many viable examples of dark matter based on UV complete models that 
do not introduce additional heavy degrees of freedom: these include keV-scale sterile neutrinos, and light scalar and vector 
fields that are populated via the freeze-in mechanism and/or vacuum misalignment (see {\em e.g.} 
\cite{Dodelson:1993je,McDonald:2001vt,Piazza:2010ye} for some early ideas). Therefore, the existence of dark matter cannot unambiguously be 
used as a pointer to new UV physics, any more than the existence of neutrino mass can be used in the same way. This may change if, for example, very energetic products of dark 
matter annihilation or decay are discovered that would be hard to accommodate within the light NP paradigm. Nonetheless, the observation that various scenarios currently remain 
open and experimentally testable has motivated the analysis in this paper, namely surveying the possible implications of light NP for precision measurements.

We have explored the sensitivity of several classes of precision observables to UV-complete models of light NP. While it is common to automatically interpret precision measurements 
in terms of generic UV new physics scenarios, we have emphasized that many of these observables are often most simply considered within models of weakly-coupled hidden sectors.
Operationally, the measurement of any deviation from a SM prediction will necessarily lead to a lengthy process to uncover its origin.
Even conclusive evidence for a new phenomenon (neutrino oscillation, dark matter) still entails further work to discriminate between viable interpretations 
involving light NP and physics at the EW scale and above. Currently, this process is underway for the measurement of $g-2$ for muons, where 
both light NP and EW scale phenomena may in principle be causing the discrepancy with SM predictions. Another existing hint of a deviation, in hadronic flavor physics (semileptonic 
B decays at LHCb), points instead to models of NP which necessarily involve new degrees of freedom at the EW scale or above. The exercise performed in this paper generalizes these 
examples to a broader set of precision observables and broader classes of models for light NP. 
In particular, we have found that the neutrino portal allows for a description of many observables in the leptonic sector, while observation of a nonzero electron EDM or related 
leptonic $CP$-violating observables would point to NP at or above the electroweak scale. In addition, we observed that several classes of observables that intrinsically involve 
hadronic flavor violation, baryon number violation, or changes to the charged currents (and thus electroweak symmetry breaking), seemingly allow an unambiguous interpretation in 
term of new short-distance physics.

\begin{acknowledgments}
The work of  M.L., M.P. and A.R. is supported 
in part by NSERC, Canada, and research at the Perimeter Institute is supported in part by the Government 
of Canada through NSERC and by the Province of Ontario through MEDT. 
\end{acknowledgments}

\bibliography{lightNP_bib}

\end{document}